\documentstyle[aps,epsfig,preprint]{revtex}

\begin{document}
\tolerance=100000
\author{ Zafar Ahmed \\
Nuclear Physics Division, Bhabha Atomic Research Centre \\
Trombay, Bombay 400 085, India \\ zahmed@apsara.barc.ernet.in }
\title
{A practice problem for Schr{\"o}dinger transmission}
\date{\today}
\maketitle
\begin{abstract}
We present the case of quantal transmission through a smooth, single-piece
exponential potential, $V(x)=-V_0 e^{x/a}$, in contrast to the piece-wise
continuous potentials, as a pedagogic model to demonstrate the analytic
extraction of transmission (reflection) co-efficient.
\end{abstract}
\vspace {.2 in}
\par The Schr{\"o}dinger transmission through a one-dimensional potential
well/barrier constitutes an essential topic in the courses and text-books [1,2]
on quantum mechanics. The main interest is to find the probability of transmission,
$T(E)$, and that of reflection, $R(E)$, when a particle is incident with an energy $E$ at the
potential from one side. For a real potential these two probanilities are such
that $T(E)+R(E)=1$, ensuring the conservation of the incident flux of the particles.
Due to simplicity of treatment, usually a rectangilar well/barrier or a sharp
step-potentials are included as examples. These potentials are both piece-wise constant
and piece-wise continuous, the former is a three-piece and the latter is a two-piece function.
\par For the case of three-piece potentials : $V(|x|\ge a)=0$
and $V(-a \le x \le a)= f(x)$, where the Schr{\"o}dinger equation does not possess
analytic form, the numerical computation of $T(E)$ is performed by letting
$\Psi(x\le -a)=A e^{ikx}+B e^{-ikx}$, $\Psi(-a\le x\le a)=C u(x)+D v(x)$
and $\Psi(x\ge a)=F e^{ikx}$ [3]. The functions $u(x)$ and $v(x)$ denote two
linearly indedependent solutions of the Schr{\"o}dinger equation with initial
values as $u(0)=1,u^\prime(0)=0; v(0)=0, v^{\prime}(0)=1$, for numerical integration
of the Schr{\"o}dinger from $x=0$ upto $x=\pm a$. Here the prime denotes differentiation
with respect to $x$.
For the case of a rectangular well/barrier of depth/height, $V_0$, these functions
are $\cos\kappa x$ and ${\sin\kappa x \over \kappa}$, where $\kappa =
\sqrt{2m(E\pm V_0)}/\hbar$. The
\begin{figure}[h]
\begin{center}
\psfig{figure=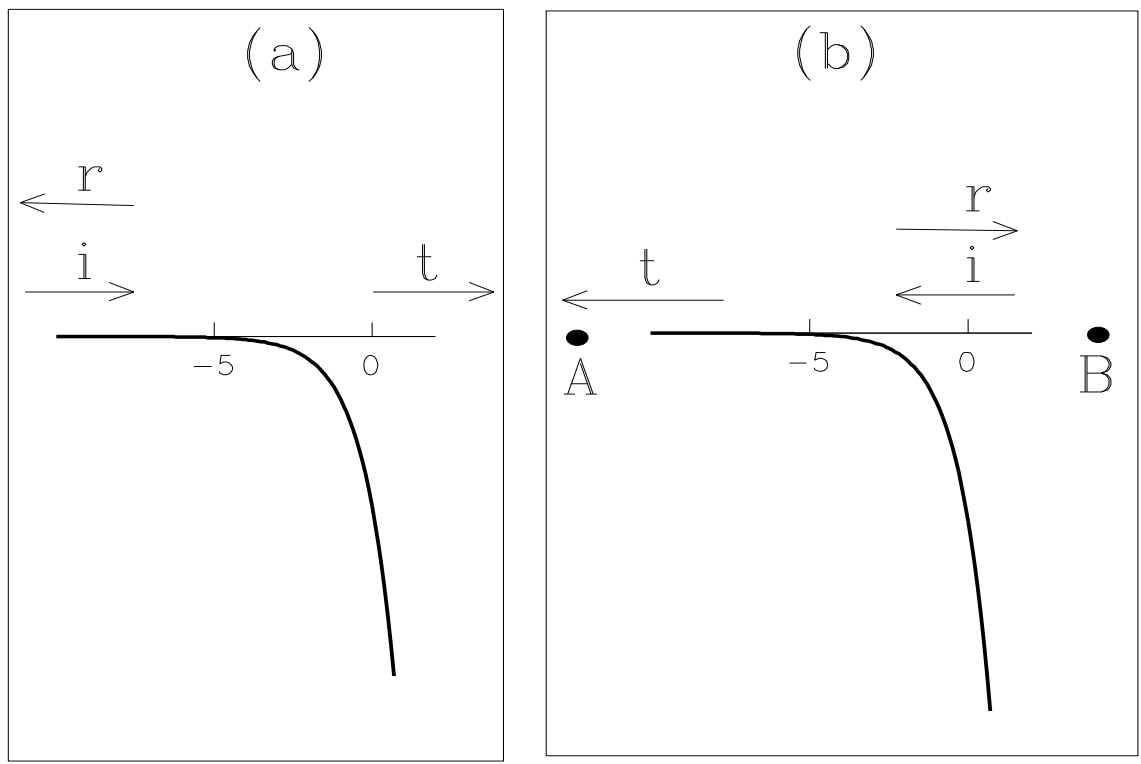,width=15.cm,height=10.cm}\\
\end{center}
\caption{Depiction of the exponential barrier Eq. (1), when $V_0=1$ and $a=1$
in arbitrary units. Here i,t,r denote the direction of incident, reflected and
transmitted waves : (a) when the particle is incident from left, (b) when the
particle is incident from right. In part (b), the filled circles represent two
bodies A and B separated by a long-ranged interaction which is presently
hypothesized as exponential potential as in Eq. (1).}
\end{figure}
\noindent
matching of the wavefunctions and their
derivatives at $x=\pm a$, leads to the elimination of $C$ and $D$ and we can
get the complex transmission amplitude as $t(E)=C/A=\sqrt{T(E)}e^{i\theta(E)}$
and complex reflection amplitude as $r(E)=B/A=\sqrt{R(E)}e^{i\phi(E)}.$
The energy-dependent quantities $\phi(E)$ and $\theta(E)$ are called reflection
phase-shift and transmission phase-shift, respectively. The energy derivative of the phase-shift
is interpreted as the time spent by a particle in these processes [1,2].
\par However, in  practical situations the potentials are smooth
(continuous and diferentiable everywhere) and single piece
which may or may not vanish asymptotically. These potentials are the parabolic
barrier [4], the Eckart barrier [3,4,5], the Fermi-step barrier[3,5], the Rosen-Morse
barrier [6], the Ginocchio barrier [7], the Scarf barrier [8],
the Morse barrier [9] and a potential which interpolates between Morse and Eckart
barriers [10]. The transmission co-efficients for these models have been obtained
analytically. In all such cases [3-10], a single function  is obtained as the
solution of the  Schr{\"o}dinger equation which behaves as a transmitted wave on
one side of the potential at asymptotically large distances (say, $x \rightarrow \infty $). On the other
side of the potential the same wavefunction behaves asymptotically (say,
$x \rightarrow -\infty$) as a combination of oppositely travelling incident
and reflected waves. Moreover, these asymptotic forms of the wavefunctions
may or may not be plane-waves depending on whether the potential converges
to zero or not at the asymptotic distance. This dual asymptotic behaviour of
a single wave-solution on two sides of the potential barrier which is absent
in the piece-wise constant/continuous potentials is worthwhile to demonstrate
through a simple do-able example. It will be even more valuable if one can
extract simple forms of $T(E)$ and $R(E)$ analytically.
\par In this paper, we present a smooth single-piece potential (see Fig.1) :
\begin{equation}
V(x)=-V_0 e^{x/a},
\end{equation}
which vanishes asymptotically as $x \rightarrow -\infty$ but it diverges to
$-\infty$ as $x \rightarrow \infty$. We feel that the attractive exponential
potential (1) presents a simple instance containing the essential features of
analytic extraction of transmission co-efficient for a one piece smooth
potential.
\par In the following we proceed to find $T(E)$ and discuss various aspects of
Schr{\"o}dinger transmission arising thereby. The time-independent Schr{\"o}dinger
equation with (1) is written as
\begin{equation}
{d^2 \psi(x)\over dx^2}+ {2m \over \hbar^2} [E+V_0 e^{x/a}]\psi(x)=0,
\end{equation}
where $m, E$ and $\hbar(={h \over 2\pi})$ denote the mass and energy of the incident
particle and $h$ is the Planck-constant. The time-dependent wave function is written
as
\begin{equation}
\Psi(x,t)=\psi(x) e^{iEt/\hbar}.
\end{equation}
Let us substitute $z=p e^{x \over 2a}$ in Eq. (2) and write
\begin{equation}
{d^2\psi \over dz^2}+{1 \over z} {d\psi \over dz}+(1-{i^2q^2 \over z^2})\psi=0,
\end{equation}
where $p= \sqrt{8mV_0 a^2 \over \hbar^2}$ and $q=\sqrt{8mE a^2 \over \hbar^2}=2ka$.
The transformed equation is the standard cylindrical Bessel equation [11]
whose linearly independent solutions for any positive energy are the Bessel
functions : $J_{iq}(z)$  and $J_{-iq}(z)$
or the Henkel functions : $H^{(1)}_{iq}(z)$ and $H^{(2)}_{iq}(z)$. Noting
the asymptotic property that
\begin{equation}
H^{(1),(2)}_{iq}(z) \sim \sqrt{2/(\pi z)}e^{\pm i(z-iq\pi/2-\pi/4)},~~
\mbox {when}~ |z| \rightarrow \infty,
\end{equation}
as $z(=p e^{x\over 2a})$ is large for large positive values of $x$, we admit
\begin{mathletters}
\begin{equation}
\Psi(x,t)= H^{(1)}_{iq}(p e^{x \over 2a})~ e^{iEt/\hbar},
\end{equation}
\begin{equation}
\approx \sqrt{2 \over \pi p} e^{-i\pi/4} e^{q\pi/2} e^{-{x \over 4a}}
\exp [i(pe^{x \over 2a} -E t/\hbar)],~~\mbox{when}~x\rightarrow \infty,
\end{equation}
\begin{equation}
\equiv \sqrt{2 \over \pi p} e^{-i\pi/4} e^{q\pi/2}  e^{-{x \over 4a}} \exp[i\Phi(x)].
\end{equation}
\end{mathletters}
This represents a wave which is exponentially decaying yet travelling towards
$x \rightarrow \infty$. The condition of constancy of the phase i.e.,
${d \Phi \over dt}=0$ yields positive phase velocity : ${dx \over dt}
={2Ea \over \hbar p} e^{-{x\over 2a}}$ which is positive definite
and does not change sign and hence the direction of the wave is unchanged.
Thus Eq. (6a) represents a transmitted
wavefunction, $\Psi_t$. The other degenerate and linearly
independent solution of the Schr{\"o}dinger equation (2),
$H^{(2)}_{iq}(z)$
has been dropped. It is instructive to note that a linear combination of
these solutions can not produce the boundary condition of a transmitted wave,
traveling towards $x \rightarrow \infty$.
\par Next, we note valuable identities :
\begin{equation}
H^{(1)}_\nu(z)=i[e^{-i\pi\nu} J_{\nu}(z)-J_{-\nu}(z)]/\sin(\nu \pi), \mbox{and}
H^{(2)}_\nu(z)=i[J_{\nu}(z)-e^{i \pi \nu}J_{-\nu}(z)]/\sin(\nu \pi).
\end{equation}
We can also write Eq. (6a) as
\begin{equation}
\Psi(x,t)={1 \over \sinh  q\pi} \left [ e^{q\pi} J_{iq}({p\over 2}e^{-x \over 2a})
-J_{-iq}({p\over 2}e^{-x \over 2a}) \right] e^{-iEt \over \hbar}.
\end{equation}
The asymptotic behaviour of Bessel function for small $z$-values is given by
[11]
\begin{equation}
J_{\pm \nu}(z)\approx {(z/2)^{\pm \nu} \over \Gamma[1\pm \nu]}.
\end{equation}
Consequently, we can write by combining Eqs. (8) and (9)
\begin{equation}
\Psi(x,t)\approx {e^{q\pi}(p/2)^{iq} \over \sinh q\pi \Gamma[1+iq]}
e^{i(kx-Et/\hbar)}-
{(p/2)^{-iq} \over \sinh  q\pi \Gamma[1-iq]}e^{-i(kx+E t/\hbar)},~~
x \rightarrow -\infty.
\end{equation}
When $x \rightarrow -\infty$, $z$ tends to a very small value, for the result
(9) to hold.
The right side of the above equation represents a linear combination of two
oppositely travelling plane waves. The phase-velocities of the first and
second part are $2E/(\hbar k)$ and $-2E/(\hbar k)$, respectively. Since the
transmitted wave has been identified to have positive phase-velocity travelling
from left to right so as to leave the barrier, the first part represents the
incident wave, $\Psi_i$ travelling from left to right impinging upon the barrier.
Similarly, the second part represents a reflected wave, $\Psi_r$, travelling
from right to left. It is instructive to note here that the asymptotic form of
$\Psi(x,t)$ as $x \rightarrow -\infty$ is a linear combination of the plane-waves
since the potential (1) actually vanishes at a large distance on the left side of
the barrier (see Fig. 1).
\par The wave-function $\Psi(x,t)$ in Eqs. (8) and (10) are nothing but
exact and asymptotic forms of the sum of incident wave, $\Psi_i$, and reflected
wave, $\Psi_r$, wavefunctions, respectively. Having found the incident, reflected
and transmitted wavefunctions, we now find the flux,
\begin{equation}
{\cal J}={\hbar \over 2im}\left (\Psi^\ast(x,t) {d\Psi(x,t) \over dx}
-\Psi(x,t){d\Psi^\ast(x,t) \over dx}\right)
\end{equation}
arising due to $\Psi_i, \Psi_r,\Psi_t$. Yet another instructive point is in order
here. The flux can also be written as a Wronskian : ${\cal J}={\hbar \over 2im}
W[\Psi^\ast,\Psi]$. Since $\Psi$ and $\Psi^\ast $ are the linearly independent
solution of (2), the Wronskian, $W[\Psi^\ast,\Psi]$,  will be constant (independent of $x$) of motion (see e.g.,
p. 48 in [2]). Therefore, if the (approximate) asymptotic forms of $\Psi_i, \Psi_r,
\Psi_t$ in Eqs. (6b) and (10) are used they will yield the same value of flux $S$
as the exact wavefunctions will do. One can check this point for the present problem
by recalling the Wronskians [11],
\begin{mathletters}
\begin{equation}
W[J_{+\nu}(z),J_{-\nu}(z)]=-{2\sin\nu \pi \over \pi z},~
W[H^{(1)}_{\nu}(z),H^{(2)}(z)_{\nu}]=-{4i \over \pi z}~
\end{equation}
and
\begin{equation}
[H^{(1)}_{iq}(z)]^\ast=- e^{\pi q} H^{(2)}_{iq} (z).
\end{equation}
\end{mathletters}
Thus, the various fluxes work out to be
\begin{equation}
{\cal J}_i={\hbar k e^{2\pi q} \over \pi m q \sinh\pi q},~
{\cal J}_r={\hbar k \over \pi m q \sinh\pi q},~{\cal J}_t={\hbar e^{\pi q}
\over \pi m a}.
\end{equation}
In obtaining the above results we have used the formula : $\Gamma(z)\Gamma(1-z)
={\pi \over \sin \pi z}$ [11].
Finally, the transmission and reflection co-efficients are found as
\begin{equation}
T(E)={{\cal J}_t \over {\cal J}_i}=[1-e^{-2\pi q}],~R(E)={{\cal J}_r
\over {\cal J}_i}=e^{-2\pi q},~ q=\sqrt {E\over \Delta},~~\Delta={\hbar^2
\over 8ma^2}.
\end{equation}
These co-efficients satisfy $T(E)+R(E)=1$. Let us find the special values of
$T(E)$ and $R(E)$ for consistency. We get $T(\infty)=1, R(\infty)=0$; $T(0)=0,
R(0)=1$ which are physically feasible. Interestingly, these co-efficients
(14) do not depend up on the parameter, $V_0$, and rightly so. Since any finite
positive value of $V_0$ is expressible as $V_0=e^{\pm b/a}$, and it would merely
result in shifting the whole potential on the $x-$axis by a distance $\pm b$ :
$V(x)=-e^{(x \mp b)/a}$.
\par In mathematical terms, when $\hbar \rightarrow 0 \Rightarrow q\rightarrow
\infty$, the classical limits are obtained as $T_{classical}(E)=1$,
and $R_{classical}(E)=0$ from Eq. (14). However, practically the classical
limit is reached when an effective parameter, for example $\Delta$ (14), becomes
extremely small. This in turn is materialized when  the tunneling
system becomes massive or/and a length scale found in the interaction (e.g., $a$
here) becomes extremely large.
In the cases [3-10], where a potential offers a barrier of a height, $V_0$, the
classical limit of $T(E)$ is a step function of energy : $T_{classical}(E<V_0)=0,
T_{classical}(E\ge V_0)=1$. Since the exponential potential (1) does not offer
any barrier to an incident particle in a classical sense, we have got
$T_{classical}(E)=1$.
\par Let us divide Eqs. (10) and (6b) by the co-efficent of $e^{i(kx-Et/\hbar)}$
in (10) to re-write $\Psi(x,t)$ respectively as
\begin{mathletters}
\begin{equation}
\Psi(x,t) \approx
e^{i(kx-Et/\hbar)}- \left (
e^{-\pi q} (p/2)^{-2iq} {\Gamma[1+iq] \over \Gamma[1-iq]}\right)
e^{-i(kx+E t/\hbar)} ,~~ x \rightarrow -\infty,
\end{equation}
\begin{equation}
\Psi(x,t) \approx \left (\sqrt{2 \over \pi p} e^{-i\pi/4} (p/2)^{-iq}
\sinh q\pi \Gamma[1+iq] e^{-q\pi/2} \right ) e^{-{x \over 4a}} \exp [i(pe
^{x \over 2a}-E t/\hbar)],~~\mbox{when}~x\rightarrow \infty.
\end{equation}
\end{mathletters}
The multiplicative factor of $e^{-i(kx_+Et/\hbar)}$  in Eq. (15a) is called as
reflection amplitude, $r(E)$ and the factor written in big brackets in Eq. (15b)
is called transmission amplitude.
\begin{equation}
r(E)=e^{-q\pi}(p/2)^{-2iq}{\gamma(1+iq) \over \Gamma(1-iq)},~~
t(E)=\sqrt{2 \over \pi p} e^{-i\pi/4} e^{-\pi q/2} (p/2)^{-iq} \sinh \pi q \Gamma(1+iq).
\end{equation}
Generally, $R(E)=|r(E)|^2$, however, $T(E)=|t(E)|^2$ holds only if the potential
converges to the same asymptotic value on both the sides of the barrier.
Since here it is not the case (see Fig. 1), the expression of $T(E)$ as obtained
in Eq. (14) by using the ratio of transmitted to incident flux, is more fundamental
and one finds that $|t(E)|^2$ is proportional to (if not identical) to $T(E)$.
Let us define Arg$[(p/2)^{iq}]=\alpha$ and Arg$[\Gamma(1+iq)]=\beta$.
For the exponential potential barrier, we get
\begin{equation}
\phi(E)=-2\alpha+2\beta\pm\pi~~\mbox {and}~~\theta(E)=-\alpha+\beta-\pi/4.
\end{equation}
\par When a particle is transmitted through a potential from one side to the other,
it sees the total potential in every detail. Eventually, detailed balancing takes
place to give rise to reciprocity of transmission :  it does not matter if the
particle is incident from left of right the transmission amplitude remains the
same (invariant), i.e., $t_{left}(E)=t_{right}(E) [1,2].$ This implies the invariance
of both transmission probability i.e., $T_{left}(E)=T_{right}(E)$
and transmission phase-shift i.e., $\theta_{left}(E)=\theta_{right}(E)$.
Further, due to the conservation of flux invariance of reflection probability
also takes place, namely, $R_{left}(E)=R_{right}(E).$
\par The behaviour of $\phi-\theta$ with respect to the side of the incidence
the particle is known (p. 109 [1]) as.
\begin{equation}
[\phi_{left}(E)-\theta_{left}(E)]
=(2n+1)\pi-[\phi_{right}(E)-\theta_{right}(E)],~n=0,1,2,...
\end{equation}
Interestingly, it turns out that if the potential is real and symmetric, we
have $\phi_{left}=\phi_{right}$ [12], and hence $\phi-\theta=\pi/2$.
The present potential being non-symmetric provides an opportunity for a simple
pedagogic demonstration of the result in Eq. (18) and the invariance of $T(E),
\theta(E)$ and $R(E)$ with respect to the side from which the particle is
incident on the potential.
\par In the following, we proceed to repeat the calculations when the side of
the incidence has been changed from left to right (see Fig. 1(b)). Using the
standard relations (7) : we find that
\begin{equation}
2e^{-i\pi\nu}J_{\nu}(z)=e^{-2i\pi\nu} H^{(2)}_{\nu}(z)+H^{(1)}(z).
\end{equation}
This identity suggests that
\begin{mathletters}
\begin{equation}
\Psi(x,t)=2e^{-q\pi} J_{-iq}({p\over 2} e^{x \over 2a}) e^{-iEt/\hbar},
\end{equation}
which in turn has an analytic behaviour for $x\rightarrow -\infty$,
\begin{equation}
\Psi(x,t) \approx 2e^{-q\pi} {(p/2)^{-iq} \over \Gamma[1-iq]} e^{-i(kx+Et/\hbar)}.
\end{equation}
\end{mathletters}
This represents a wave travelling to the left after being transmitted from the
potential (see Fig. 1(b)). Using Eq.(9) and the asymptotic properties of
$H^{(1),(2)}_{\nu}$ as in Eq. (5), we find that
\begin{equation}
\Psi(x,t)\approx 2 \sqrt{2 \over \pi p} e^{i\pi/4} e^{-3\pi q/2} e^{-x \over 4a}
\left ( e^{-i({p \over 2} e^{x \over 2a}+Et/\hbar)}-i e^{-\pi q} e^{i({p \over 2}
e^{x \over 2a}-Et/\hbar)} \right ),~~\mbox {when}~~x\rightarrow \infty.
\end{equation}
In the asymptotic form of the wave function (21), the first part denotes the
incident wave, $\Psi_i$, entering from the right side of the potential
(see Fig.1(b)) and travelling to the left. The second part denotes oppositely
travelling reflected wave, $\Psi_r$. The transmitted wave, $\Psi_t$, is
represented by Eq.(19). One can recover, the results found in Eq. (14)
by evaluating the flux arising due to these wave-forms when the side of incidence
has been changed. The recovery of $T(E)$ and $R(E)$ despite the change of the
side of the incidence of the particle on the potential (see Fig. 1(b))
demonstrates their invariance with respect to the side of incidence of the
particle whether it is left or right.
\par Once again, let us divide Eqs. (19b) and (20) by the co-efficient written
outside the big bracket in Eq. (20) to get
\begin{mathletters}
\begin{equation}
\Psi(x,t) \approx \sqrt{\pi p \over 2} e^{-i\pi/4} e^{q\pi/2} {(p/2)^{-iq} \over
\Gamma[1-iq]} e^{-i(kx+Et/\hbar)}, ~~ \mbox {when}~~x\rightarrow -\infty,
\end{equation}
\begin{equation}
\Psi(x,t)\approx  e^{-x \over 4a} \left (
e^{-i({p \over 2} e^{x \over 2a}+Et/\hbar)}-i e^{-\pi q} e^{i({p \over 2}
e^{x \over 2a}-Et/\hbar)} \right ),~~\mbox {when}~~x\rightarrow \infty.
\end{equation}
\end{mathletters}
The amplitudes can be extracted as earlier :
\begin{equation}
r_{right}(E)=-i e^{-\pi q}~~\mbox {and}~~ t(E)=\sqrt{\pi p \over 2} e^{-i\pi/4}
e^{q\pi/2} {(p/2)^{-iq} \over \Gamma[1-iq]}.
\end{equation}
Further, we find the phase-shifts as
\begin{equation}
\phi_{right}(E)=3\pi/2 ~~\mbox {and}~~\theta_{right}(E)=-\alpha+\beta-\pi/4
\end{equation}
to compare with phase-shifts given in Eq. (17), and
this completes the demonstration of the general results given in Eq. (18).
\par It is further relevant to know that when the potential is complex,
the reciprocity of the transmission amplitude with respect to the side of the
incident particle still holds. However, reflection and absorption amplitudes and co-efficients
depend on the side (left/right) of incident particle, if only, the total
complex potential is non-symmetric, namely $V_{complex}(-x) \ne V_{complex}(x)$
[13].
\par The utility of the Schr{\"o}dinger transmission in many physical processes
can be appreciated as follows. Look at one body, $A$, at rest in Fig. 1(b)
and let other body, $B$, come from a side with an energy, $E$. Assume that
the interactions between $A$ and $B$ are of two types : short-ranged and
long-ranged.
Thus, the interaction barrier (e.g. Eq.1(b)) due to long-ranged interaction has
to be overcome by the body $B$ in order to appear on the other side to have any
worthwhile contact/reaction with $A$, characterized by their short-ranged
interaction. The probability of the appearance of the body, $B$, on the other
side is clearly proportional to the probability of transmission through the potential
barrier. This suggests that the reaction rates of $A$ and $B$, would depend on the
transmission probability, $T(E)$, apart from other factors.
\par In fact, barrier
penetration method of determining the fusion rates (see [10] and Refs. therein)
of nuclei is one such instance where fusion rates are determined by finding the
penetrability $(T(E))$ of the inter-nuclear potential barrier formed due to
short-ranged nuclear attraction and Coulomb plus centrifugal repulsion.
\par In other kind of instances, a bound system (e.g. $\alpha$-particles
in a nucleus [14] or electrons of a metal in the field emission [15])
faces a barrier and the emission rates are then determined by the transmission
co-efficient of the barrier.
\par The analytically solvable, smooth and single-piece potential barriers
mentioned in the introduction find applications
in several physical processes. The extraction of transmission co-efficient
for these barriers involves higher order functions. In this regard, the exponential
barrier (1) presented here, despite being simple allows one to have the actual
experience of extracting transmission (reflection) amplitude/co-efficent from a
smooth, single-piece potential analytically. This problem may be practised right
after doing rectangular well/barrier and before attempting the ``higher-order''
problems found in Refs. [3-10]. The exponential potential model also illustrates,
with simplicity, several other aspects of Schr{\"o}dinger transmission from one-dimensional,
real potentials.
\section*{Acknowledgements}
I would like to thank Dr Sudhir. R. Jain and Dr Abhas Mitra for critical
reading of the manuscript.
\section*{references}
\begin{enumerate}
\item A. Messiah, {\it Quantum Mechanics} (North-Holland Publishing. Co.,
Amsterdam, 1962) vol. I, pp. 91-114.
\item E. Merzbacher, {\it Quantum Mechanics} (John Wiley \& Sons, New York,
1970) Ch. 6, pp. 80-114
\item S. Fl{\"u}gge, {\it Practical Quantum Mechanics} (Springer-Verlag, Berlin,
1971) vol. 1, Prob. No. 37 \& 39.
\item D. Rapp, {\it Quantum Mechanics} (Holt,Rinhart \& Wilson Inc., New York,
1970), Part III, Ch. 8, pp. 125-200; G. Barton, ``Quantum mechanics of inverted
harmonic oscillator potential'', Ann. Phys. (N.Y.) {\bf 166}, 322 (1986);
Z. Ahmed, ``Tunneling through asymmetric parabolic barriers'', J. Phys. A: Math.
Gen. (1997)) 3115-3116.
\item L.D. Landau and E. M. Lifshitz, {\it Quantum Mechanics} (Pergamon Press,
London,1958).
\item P.M. Morse and H. Feshbach, {\it Methods of Theoretical Physics}
(McGraw-Hill Book Company, Ltd., New York, 1953) pp.1650-1660.
\item J.N. Ginocchio, ``A class of exactly solvable 1: one-dimensional
Schr{\"o}dinger equation'', Ann. Phys. {\bf 152}, 203 (1984);
B.Sahu, S.K. Agarwalla and C.S. Shastry, ``Above-barrier resonances :
Analytic expression for energy and width'', J. Phys. A: Math. Gen. {\bf 35},
4349-4358 (2002).
\item A. Khare and U. P. Sukhatme, ``Scattering amplitudes for supersymmetric
shape-invariant potential by operator method'', J. Phys. A: Math. Gen. {\bf 21}
L501-L508 (1988).
\item Z.Ahmed,``Tunneling through the Morse barrier'', Phys. Lett. A {\bf 157},
1-5, (1991).
\item Z.Ahmed, ``Tunneling through a one dimensional potential barrier'',
Phys. Rev. A, {\bf 47} 4761-4757 (1993).
\item G.B. Arfken and H.J. Weber, {\it Mathematical Methods for Physicists},
(academic Press Inc., San Diego,1995) pp. 627-664;
M. Abramowitz and A. S. Stegun, {\it Handbook of Mathematical Functions} (Dover
Publications, Inc., New York, 1970) pp.358-365.
\item B.F. Buxton and M.V. Berry, ``Bloch wave degeneracies in systematic high
energy electron diffraction'',
Phil. Trans. R. Soc. London. Ser. A {\bf 282}, 485-525 (1976) (see p.534);
G. Barton, ``Levinsion theorem in one dimension : Heuristics'', J. Phys. A
{\bf 18} 497 (1985); Y. Nogami and C.Y. Ross, ``Sacattering from a non-symmetric
potential in one dimension : as a coupled channel problem'', Am . J. Phys., {\bf 64},
923 (1996).
\item Z. Ahmed, ``Schr{\"o}dinger transmission through one-dimensional
complex potentials'', Phys. Rev. A {\bf 64} 042716-(1-4) (2001).
\item e.g., H. A. Enge,{\it Introduction to Nuclear Physics} (Addison Wesley
Publishing Co., Reading, Mass., 1966), pp. 274-295.
\item e.g., A.S. Davydov, {\it Quantum Mechanics} (Pergamon Press, New York,
1965).
\end{enumerate}
\end{document}